\documentclass[reprint, aps, twoside, floatfix, nofootinbib, longbibliography]{revtex4-2}
\usepackage{amsmath, amssymb}
\usepackage{graphicx}
\usepackage[usenames, dvipsnames]{color}
\usepackage{hyperref}
\hypersetup{colorlinks=true, linkcolor = blue, citecolor = blue, urlcolor = blue}
\usepackage{dcolumn}

\newcommand\he[1]{\multicolumn{1}c{#1}} 

\newcommand\hel[1]{\multicolumn{1}l{#1}} 

\newcommand{\opS}{\hat{\mathrm{S}}}

\newcommand{\opH}{\hat{\mathcal{H}}}

\newcommand{\cC}{\mathcal{C}}

\newcommand{\res}{\mathrm{r}}
\newcommand{\dd}{\mathrm{d}}
\newcommand{\ee}{\mathrm{e}}
\newcommand{\ii}{\mathrm{i}}
\newcommand{\pdt}{\partial_t}

\newcommand{\vacqket}{|\varnothing_q\rangle}
\newcommand{\vacqbra}{\langle\varnothing_q|}
\newcommand{\omegaa}{\omega_\mathrm{a}}
\newcommand{\omegac}{\omega_\mathrm{c}}

\newcommand{\omegar}{\omega_\mathrm{r}}

\newcommand{\Nres}{N}
\everymath{\displaystyle}
\begin{document}
\title{Single-photon switch controlled by a qubit embedded in an engineered electromagnetic environment}
\author{E.~V.~Stolyarov}
\email{eugenestolyarov@gmail.com}
\affiliation{Institute of Physics of the National Academy of Sciences of Ukraine, pr. Nauky 46, 03028 Kyiv, Ukraine}
\begin{abstract}
 A single-photon switch is an important element for the building of scalable quantum networks.
 In this paper, we propose a feasible scheme for efficient single-photon switching.
 The proposed switch is controlled by a state of a qubit formed by the pair of the lowest levels of a three-level system (qutrit) coupled to a resonator.
 This resonator-qutrit system comprises a switching unit of the considered setup.
 For suppression of the Purcell relaxation of the control qubit, the switching unit is embedded into a coupled-resonator array serving as an engineered electromagnetic environment with a band gap on a qubit transition frequency.
 We discuss the possible implementation of the considered single-photon switch on the microwave circuit QED architecture.
 We demonstrate that high switching contrasts can be attained for the parameters achievable for the state-of-the-art superconducting circuit QED setups.
\end{abstract}
\setcounter{page}{1}
\maketitle
\section{Introduction} \label{sec:intro}
A quantum network is an essential ingredient necessary for the realization of scalable systems for quantum information processing (QIP) \cite{kimble2008}.
It is built of a set of nodes, where quantum information is processed and/or stored, interconnected via quantum channels, where flying qubits propagate transferring information between remote quantum nodes \cite{ritter2012}.
Photons are considered as a prime candidate for the role of flying qubits due to the ultimate propagation speed and the ability to retain the coherence over the large distances \cite{northup2014}.
Precise and rapid control of photon propagation in quantum networks is requisite for the efficient operation of quantum networks.
In this regard, various devices aimed to manipulate the transport of photons, such as quantum switches \cite{bermel2006,lzhou2008,jqliao2009,xia2013,parkins2014,yan2016,ytzhu2019,agham2019} and routers \cite{zhou2013,yan2014,lu2014,ahumada2019,zhu2019,poudyal2020}, photonic valves \cite{masc2014}, diodes \cite{roy2010,wbyan2018}, and transistors \cite{chang2007,neumeier2013,kyr2016}, were proposed.

A single-photon switch is a system that coherently controls the photonic transport on a level of individual quanta.
A switch interconnects different quantum channels and represents an important component (node) of quantum networks, which motivates the studies of various schemes for switching and routing.
Besides a plethora of theoretical proposals \cite{bermel2006,lzhou2008,jqliao2009,zhou2013,lu2014,yan2016,zhu2019,ytzhu2019,agham2019,ahumada2019}, a number of experimental demonstrations of various schemes of single-photon switches and routers operating in both microwave and optical domains were reported \cite{dayan2008,aoki2009,hoi2011,oshea2013,papon2019}.

Waveguide QED structures, such as optical nanofibers \cite{nieddu2016,*nayak2018}, photonic-crystal waveguides \cite{lod2015}, or coplanar microwave transmission lines \cite{gu2017}, can serve as quantum channels providing robust transmission of photons.
In waveguides, light is transversely confined, which gives rise to light-emitter interaction enhancement and pronounced interference between the incident and scattered fields. 
It was demonstrated that an individual quantum emitter embedded in a one-dimensional waveguide can act as a tunable scatterer for an incident photon \cite{jtshen2005,chyan2011,chu2013}.
By varying the strength \cite{zhou2008,jqliao2009,chyan2014} and phase \cite{yang2018} of light-emitter coupling or utilizing a control field (classical \cite{bermel2006,zhou2013,lu2014,ytzhu2019,ahumada2019,agham2019} or quantum \cite{kolchin2011,wbyan2014,*yan2015epl}), one can achieve either complete transmission or reflection of an incident (probe) photon.
This feature is used for the implementation of optical switches and routers \cite{hoi2011,oshea2013,aoki2009}.

\begin{figure*}[t!] 
	\centering
	\includegraphics{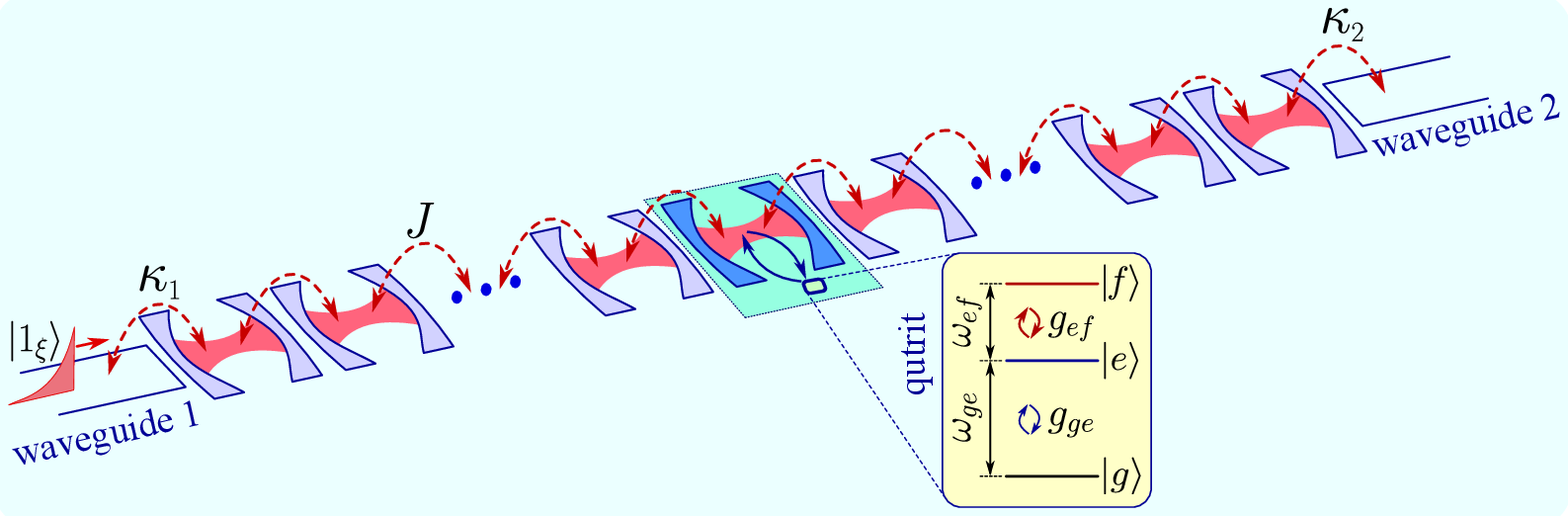}
	\caption{Scheme of the single-photon switch under analysis.
		A coupled-resonator array (CRA) is coupled on both sides to semi-infinite waveguides (marked with indices 1 and 2).
		A switching unit, highlighted by a shaded area, is composed of a resonator coupled to a 3LS (qutrit).
		The level structure of the 3LS is shown in the inset.
		\label{fig:fig_1}}
\end{figure*}

In the paper, we propose a scheme of an efficient single-photon switch based on a waveguide QED system which can be realized on a microwave superconducting circuit QED (cQED) hardware platform.
In the scheme we consider, the pair of semi-infinite waveguides is coupled to ends of a coupled-resonator array (CRA).
One of the resonators composing the CRA is coupled to a three-level system (3LS) implemented by a Josephson-junction artificial atom.
This resonator-qutrit system works as an active (switching) unit in the considered scheme. 
The two lowest states of a 3LS constitute a qubit the state of which controls whether the system transmits or reflects the input photon.
Thus, there is no need in the continuous classical drive to switch the system between the reflective and transmissive states, which is required in various proposals of single-photon switches \cite{bermel2006,xia2013,zhou2013,lu2014,agham2019,ahumada2019}.
In the considered setup, one requires only short classical control pulses for preparation of the qubit state \cite{motzoi2009,*gamb2011}.
Moreover, recent theoretical \cite{mcdermott2014,*liebermann2016} and experimental studies \cite{leonard2019} suggest that one can use single-flux quantum pulses for this purpose.
Such an approach allows one to integrate the control electronics along with the resonators and artificial atoms on a single chip, which reduces the length of interconnects and brings most of the setup components into the cryogenic stage.

In the considered scheme, the CRA represents an engineered electromagnetic environment with a band gap.
The frequency of the controlling qubit is tuned to fall within that band gap, which inhibits the Purcell relaxation of the qubit and improves the performance of the switch.

We provide a fully quantum-mechanical description of the single-photon wave-packet transport in the system under consideration.
The dependence of the switching contrast on the system parameters is studied.
A set of parameters of the system providing the maximal switching contrast is determined.

The paper is organized as follows.
In Sec.~\ref{sec:setup} we describe the scheme, the principle of operation, and the possible cQED implementation of the proposed single-photon switch.
The model Hamiltonian of the studied system is given as well.
In Sec.~\ref{sec:transp} we derive the effective Hamiltonian of the system and use it to describe a single-photon transport.
The results of calculations of the dependence of a switching contrast on the system parameters are demonstrated in Section~\ref{sec:res}.
In Sec. \ref{sec:concl} we discuss possible extensions and applications of the considered switch and summarize the results.
Derivations of various equations of motion used in the main text are presented in Appendix~\ref{sec:appa}.
The definition of the parameter characterizing the photon spectrum modification after traversing the switch is given in Appendix~\ref{sec:spec}.

\section{Setup} \label{sec:setup}

\subsection{Scheme and operational principle} \label{sec:princ}

We consider a realization of the single-photon switch consisting of an array (chain) of an odd number $\textstyle N_\mathrm{res}=2\Nres+1$ of optical resonators.
The terminal resonators of an array are coupled to semi-infinite one-dimensional optical waveguides marked with indices 1 and 2.
In what follows, we assume that the first waveguide acts as an input dispatching the ingoing single-photon wave packet to the CRA, while the second waveguide acts as an output channeling the scattered (transmitted) photon.
The central resonator of the array is coupled to a 3LS.
In practice, the latter is represented by a superconducting artificial atom.
All resonators in the array, apart from the central resonator, have identical frequencies $\textstyle \omegar$.
The central resonator has frequency $\textstyle \omegac$.
Each resonator is coupled to its nearest neighbors with strength $\textstyle J$.
The schematic of the considered setup is presented in Fig.~\ref{fig:fig_1}.

First, let us elucidate the principle of operation of the proposed single-photon switch.
The switching unit, which controls the photon transport in the setup we consider, consists of a resonator coupled to a 3LS or qutrit.
We use the conventional notation for the qutrit eigenstates, where $\textstyle |g\rangle$ stands for the ground state, and $\textstyle |e\rangle$ and $\textstyle |f\rangle$ are excited states.
The eigenstates form a ladder configuration implying that only $\textstyle |g\rangle \leftrightarrow |e\rangle$ and $\textstyle |e\rangle \leftrightarrow |f\rangle$ transitions are allowed.
The transition frequency $\textstyle \omega_{ef}$ between $\textstyle |e\rangle$ and $|f\rangle$ levels is tuned in resonance with the resonator frequency $\omegac$.
In contrast, the transition frequency $\omega_{ge}$ between $|g\rangle$ and $\textstyle |e\rangle$ states is strongly detuned from the resonator frequency, which inhibits the excitation exchange between the $\textstyle |g\rangle \leftrightarrow |e\rangle$ transition and the resonator mode.
Such an interaction regime between the resonator mode and the $\textstyle |g\rangle \leftrightarrow |e\rangle$ transition is referred to as the \emph{dispersive coupling} regime \cite{blais2004}.

\begin{figure}[t!] 
	\centering
	\includegraphics{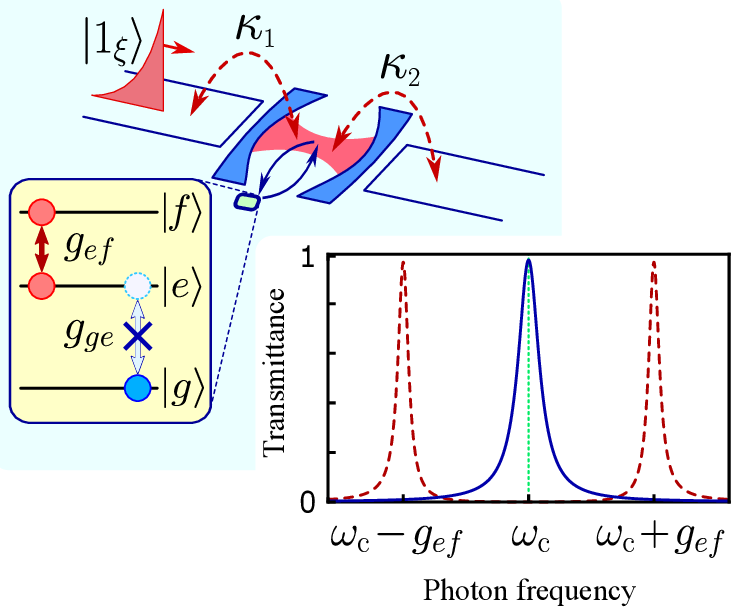}
	\caption{Scheme of the simplified version of the single-photon switch.
		The plot shows the dependence of the single-photon transmittance on the ingoing photon frequency for the uncoupled resonator (solid line) and the resonator coupled to a 2LS with strength $\textstyle g_{ef}$ (dashed line).
		The transmittance exhibits the maximum on the resonator frequency $\textstyle \omegac$ for the uncoupled resonator and the minimum for the resonator coupled to a 2LS.
		\label{fig:fig_2}} 
\end{figure}

Now, let us qualitatively explain how the resonator-qutrit system provides control over the photon scattering.
For this purpose, we consider a simplified version of the single-photon switch, which is represented by the resonator-qutrit system directly coupled to a pair of semi-infinite one-dimensional waveguides acting as input and output. The scheme of this setup is demonstrated in Fig.~\ref{fig:fig_2}.
In such a system, the transmission of itinerant photons from the input waveguide to the output waveguide can be controlled by manipulating the state of the qubit encoded by the pair of the lowest states of the qutrit, namely, $\textstyle |g\rangle$ and $\textstyle |e\rangle$.
When one prepares the qubit in the ground state $\textstyle |g\rangle$, the resonator-qutrit system acts effectively as just a resonator alone, since the interaction between the resonator mode and $\textstyle |g\rangle \leftrightarrow |e\rangle$ transition is dispersive.\footnote{Note that the frequency of this ``effective" resonator is slightly shifted compared to the frequency of the ``bare" (uncoupled) resonator.
This shift is induced by the dispersive interaction with the $\textstyle |g\rangle \leftrightarrow |e\rangle$ transition of the qutrit (see details in  Sec.~\ref{sec:ham_eff}).}
When the control qubit is prepared in the state $\textstyle |e\rangle$, the transitions between $\textstyle |e\rangle$ and $\textstyle |f\rangle$ states can occur due to excitation exchange with the resonator, while the transition from the state $\textstyle |e\rangle$ to the ground state $|g\rangle$ is inhibited due to the dispersive interaction with the resonator.
In this case, the resonator-qutrit system acts similarly to a resonator coupled to a two-level system (2LS) formed by states $|e\rangle$ and $|f\rangle$.
The single-photon transmission spectrum of such a resonator-2LS system exhibits a ``dip" on the resonator frequency in contrast to the transmittance maximum in the case of an uncoupled resonator.
This effect is referred to as the dipole induced reflection (DIR)\cite{auff2007}.\footnote{Here we consider the setup featuring the direct coupling of a resonator to a pair of semi-infinite waveguides, as shown in Fig.~\ref{fig:fig_2}.
In the case of a resonator side-coupled to a single bi-directional waveguide, one encounters the related effect referred to as the dipole induced transparency (DIT). In the latter arrangement, the single-photon transmission on the resonator frequency has a minimum for the case of an uncoupled resonator and a maximum when a resonator is coupled to a 2LS \cite{waks2006}.}

However, due to the Purcell effect \cite{sete2014}, the excited state $\textstyle |e\rangle$ relaxes to the ground state $\textstyle |g\rangle$, which deteriorates the performance of the waveguide-resonator-qutrit switch described above and illustrated in Fig.~\ref{fig:fig_2}.
To remedy this limitation and improve the switching efficiency, instead of coupling the resonator-qutrit system directly to the waveguides (as shown in Fig.~\ref{fig:fig_2}), we embed the former into an engineered electromagnetic environment with a band gap, where photons can not propagate.
The CRA can act as such an environment.
Assuming that $\textstyle \omegac\approx\omegar$, the CRA composed of $\textstyle N_\mathrm{res}$ resonators exhibits the dispersion relation $\textstyle \mathcal{E}_n = \omegar - 2J \cos q_n$, where $\textstyle q_n = n\pi/(N_\mathrm{res}+1)$.
Thus, the CRA features a passband of width $\textstyle 4J$ centered around $\textstyle \omegar$ \cite{tan2012}.
We can harness this property and specifically design the energy levels of an artificial atom (qutrit) in a way, that the transition frequency $\textstyle \omega_{ef}$ lies within the passband of the CRA, while $\textstyle \omega_{ge}$ falls into its band gap.
In this case, the Purcell relaxation of the state $|e\rangle$ to the state $\textstyle |g\rangle$ is completely suppressed \cite{bykov1975,*yabl1987,*sajeev1990}.

\subsection{Circuit QED implementation} \label{sec:cqed}

\begin{figure}[b!] 
	\centering
	\includegraphics{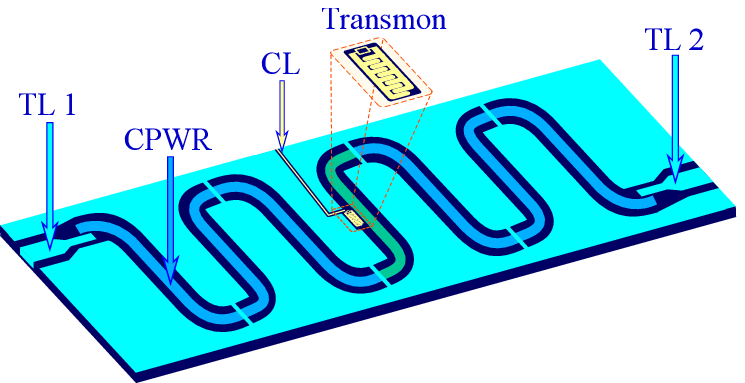}
	\caption{Schematic illustration of the potential cQED implementation of the considered single-photon switch.
		The CRA is comprised of capacitively coupled CPWRs.
		The CPWRs are arranged similarly to that in Ref.~\cite{fitz2017}.
		The state of the transmon is prepared using the control line (CL).
		Both sides of the CRA are coupled to coplanar transmission lines (TL1 and TL2).
		In this particular setup, all resonator frequencies and couplings are fixed and set on a fabrication stage.
		\label{fig:fig_3}} 
\end{figure}

Let us briefly discuss a possible experimental realization of the proposed device within the superconducting cQED architecture.
Microwave superconducting circuits provide a versatile and scalable hardware platform for the implementation of QIP devices \cite{wendin2017}.
Josephson-junction artificial atoms \cite{krantz2019} are genuinely multilevel systems offering tunable level structure and transition frequencies.

The cQED realization of the model system illustrated in Fig.~\ref{fig:fig_1} can be as follows. 
The CRA is composed of coplanar waveguide resonators (CPWR) \cite{goppl2008} interacting via capacitive couplings, which can be made either fixed or tunable.
The latter is achieved by coupling resonators via SQUIDs \cite{wulsch2016,koun2018}, which allows one to individually control the interaction strength between the resonators by changing the external flux through each SQUID loop.
However, the payoff for tunability is the increase of the setup complexity.
A pair of microwave coplanar transmission lines coupled to the terminal resonators of the CRA serves as semi-infinite one-dimensional waveguides.
The central resonator in the CRA is coupled to a transmon-type \cite{koch2007} superconducting artificial atom featuring a ladder-type structure of energy levels. This type of superconducting artificial atoms and its modifications \cite{barends2013, yuchen2014} offer high coherence times and tunable couplings, which makes it widely utilized for the building of various QIP devices \cite{gambetta2017}.
The state of the qubit is manipulated on-demand via the control line (CL) \cite{krantz2019}.
Since, in the considered scheme, the switching between the reflective and transmissive state of the device is realized via applying the \textsf{X}-gate (quantum bit-flip gate) to the qubit, the time of switching is determined by the \textsf{X}-gate time.
For transmons, the latter is typically $\textstyle \sim 20\,\textrm{ns}$ with $\geq 0.999$ fidelity \cite{barends2014}.
The cQED incarnation of the single-photon switch outlined above is feasible for the current technologies.
The sketch of this cQED setup is shown in Fig.~\ref{fig:fig_3}.
 
\subsection{Model Hamiltonian} \label{sec:ham}
The Hamiltonian describing the model system outlined in Sec.~\ref{sec:princ} reads as 
\begin{equation} \label{eq:ham}
 \opH = \opH_\res + \opH_{\res-\res} + \opH_\mathrm{s} + \opH_\mathrm{w} + \opH_\mathrm{w-r}.
\end{equation}
The first term in Eq.~\eqref{eq:ham} is the Hamiltonian of $\textstyle 2\Nres$ identical resonators with frequencies $\textstyle \omegar$:
\begin{equation} \label{eq:ham_arr}
   \opH_\res = \hbar \sum_{n=1}^{\Nres} \omegar \left(a^\dag_{-n} a_{-n} + a^\dag_n a_n\right),
\end{equation}
where $\textstyle a_n$ is the annihilation operator of a photon in the $\textstyle n$-th resonator of the CRA obeying the equal-time commutator
$\textstyle [a_n,a^\dag_{n'}] = \delta_{n,n'}$.
In what follows, the subscript $n$ is reserved for the resonator indices running sequentially from $\textstyle -N$ to $\textstyle N$.
The index $\textstyle n=0$ is attributed to the central resonator.

The second term in Eq.~\eqref{eq:ham} describes the nearest-neighbor coupling between the resonators in the array.
The Hamiltonian $\textstyle \opH_{\res-\res}$ reads
\begin{equation} \label{eq:ham_rr}
 \opH_{\res-\res} = \hbar J \sum_{n=-\Nres}^{\Nres}\left(a^\dag_n a_{n+1} + a^\dag_{n+1} a_n\right).
\end{equation}

The term $\textstyle \opH_\mathrm{s}$ describes the switching unit -- the system composed of the central resonator coupled to the ladder-configuration 3LS.
The Hamiltonian $\textstyle \opH_\mathrm{s}$ reads
\begin{equation} \label{eq:ham_sw}
 \begin{split}
  \opH_\mathrm{s} = & \, \hbar \omega_\mathrm{c} a^\dag_0 a_0 + \hbar \omega_{ge}\sigma_{ee} + \hbar (\omega_{ge} + \omega_{ef})\sigma_{ff} \\
  & \, + \hbar g_{ge}(a^\dag\sigma_{ge} + \sigma_{eg}a) + \hbar g_{ef}(a^\dag\sigma_{ef} + \sigma_{fe}a).
 \end{split}
\end{equation} 
The first term in Eq.~\eqref{eq:ham_sw} describes the central resonator.
The second and the third terms constitute the Hamiltonian of the 3LS.
The last pair of terms in Eq.~\eqref{eq:ham_sw} describe the coupling between the qutrit and the central resonator.
In Eq.~\eqref{eq:ham_sw}, we have introduced an operator $\textstyle \sigma_{kl} = |k\rangle\langle l|$, where $k,l\in\{g,e,f\}$.
This operator obeys the commutation relation as follows:
\begin{equation} \label{eq:s_comm}
 \left[\sigma_{kl},\sigma_{k'l'}\right] = \sigma_{kl'}\delta_{k'l} - \sigma_{k'l}\delta_{kl'}.
\end{equation}
Parameters $\textstyle g_{ge}$  and $\textstyle g_{ef}$ stand for the coupling strengths between the central resonator and $\textstyle |g\rangle \leftrightarrow |e\rangle$ and $\textstyle |e\rangle \leftrightarrow |f\rangle$ transitions, correspondingly.
For the transmon, these couplings are related as $\textstyle g_{ef}/g_{ge} \approx \sqrt{2}$ \cite{koch2007}.

The resonator-resonator and resonator-qutrit couplings are described within the rotating-wave approximation (RWA).
The latter is valid provided that the following criteria are satisfied
\begin{subequations}
 \begin{gather}
  \begin{split} \label{eq:rwa_freqs}
   |\omegar - \omegac| & \ll \omegar +\omegac, \\ |\omega_{ge(ef)}-\omegac| & \ll \omega_{ge(ef)} + \omegac,
  \end{split} \\ \label{eq:rwa_coupls}
  J \ll \omegar,\omegac, \quad g_{ge} \ll \omega_{ge}, \omegac, \quad g_{ef}\ll \omega_{ef}, \omegac.
 \end{gather} 
\end{subequations}
While there are experimental demonstrations of ultrastrong coupling [the  criterion~\eqref{eq:rwa_coupls} breaks down] between the resonator and the artificial atom in cQED \cite{niemczyk2010,yoshihara2016}, the use of the RWA is well justified for the range of parameters we use in the paper.

The waveguides are described by the Hamiltonian
\begin{equation} \label{eq:ham_wg}
 \opH_\mathrm{w} = \hbar \int^\infty_0 \dd\omega \omega \sum_{j=1}^2 b^\dag_{j,\omega} b_{j,\omega}.
\end{equation}
The bosonic operator $\textstyle b_{j,\omega}$ annihilates a photon with frequency $\textstyle \omega$ in the $\textstyle j$-th waveguide and obeys the commutation relation $\textstyle [b_{j,\omega}, b^\dag_{j',\omega'}] = \delta(\omega'-\omega)\delta_{j'j}$.

The Hamiltonian $\opH_\mathrm{w-r}$, which describes the couplings between the waveguides and the CRA, reads
\begin{equation} \label{eq:ham_wcra}
 \begin{split}
   \opH_\mathrm{w-r} = & \, \hbar \int^\infty_0 \dd\omega f_1(\omega) \left(b^\dag_{1,\omega} a_{-\Nres} + a^\dag_{-\Nres} b_{1,\omega}\right) \\
   & \, + \hbar \int^\infty_0 \dd\omega f_2(\omega) \left(b^\dag_{2,\omega} a_{\Nres} + a^\dag_{\Nres} b_{2,\omega}\right),
 \end{split}
\end{equation}
where $\textstyle f_{j}(\omega)$ stands for the frequency-dependent coupling of the CRA to the $\textstyle j$-th waveguide. 
The coupling between the $j$-th waveguide and the CRA gives rise to the photon exchange between them with rate $\textstyle \kappa_j = 2\pi f^2_j(\omegar)$ (see details in Appendix~\ref{sec:appa1}).
The Hamiltonian $\textstyle \opH_\mathrm{w-r}$ in Eq.~\eqref{eq:ham_wcra} is given within the RWA, assuming that $\textstyle \kappa_j\ll\omegar$.

In our model, we do not account for the dissipation processes, assuming that they occur on timescales much longer than the coherent processes in the system.
Indeed, the internal quality factor of CPWRs \cite{megr2012,bruno2015} can surpass $\textstyle 10^6$, which corresponds to the resonator dissipation rate $\textstyle \gamma_\mathrm{res}/(2\pi)\lesssim 0.01\,\mathrm{MHz}$.
The probability of photon loss in an individual resonator is determined as $\Lambda_\mathrm{res} \approx \gamma_\mathrm{res} \tau_\mathrm{res}$, with $\textstyle \tau_\mathrm{res} \sim 1/(2J)$ being a photon lifetime in a resonator.\footnote{The photon lifetime in the terminal resonators is estimated as $\textstyle \tau_{j}\sim 1/(J+\kappa_{j})$ with $\textstyle j\in\{1,2\}$.
However, assuming that $\kappa_{j}$ and $J$ are of the same order of magnitude, the estimate $\textstyle \tau_{j}\sim \tau_\mathrm{res}$ is applicable.}
Since the processes of dissipation in each resonator composing the CRA are independent, the photon loss probability in the CRA is determined as $\textstyle \Lambda_\mathrm{cra} \approx N_\mathrm{res} \Lambda_\mathrm{res}$.
Thus, the photon loss in the CRA can be neglected in the analysis of the photon transport provided that the following condition is satisfied
\begin{equation} \label{eq:cra_loss}
 \Lambda_\mathrm{cra}=\frac{N_\mathrm{res}\gamma_\mathrm{res}}{2J} \ll 1.
\end{equation}
Taking the typical parameters $\textstyle J/(2\pi) \sim 10\,\mathrm{MHz}$ and $\textstyle N_\mathrm{res} \sim 10$, one arrives at the estimate for the photon loss probability in the CRA $\Lambda_\mathrm{cra}\lesssim 0.01$.
Thus, the dissipation has a minor effect on photon transport through the CRA.
The effect of the qubit relaxation can be neglected in the case the qubit coherence time $\textstyle \tau_\mathrm{coh}$ fulfills the criterion
\begin{equation} \label{eq:qb_coh}
 \tau_\mathrm{coh} \gg \tau_\mathrm{tvl},
\end{equation}
where $\textstyle \tau_\mathrm{tvl} \approx \tau_\mathrm{p} + N_\mathrm{res}\tau_\mathrm{res}$ stands for a wave-packet travel time through the CRA.
Taking $\textstyle J/(2\pi) \sim 10\,\mathrm{MHz}$ and $\textstyle N_\mathrm{res} \sim 10$, one obtains $\textstyle \tau_\mathrm{tvl}\lesssim 1\,\mathrm{\mu s}$, while the coherence times of the modern transmons approach $\textstyle \tau_\mathrm{coh}=100\,\mu\mathrm{s}$ \cite{rig2012,jchang2013,park2020}, implying that the relaxation of the artificial atom can be neglected in the analysis of the photon transport.
Throughout the paper, we assume that the criteria \eqref{eq:cra_loss} and \eqref{eq:qb_coh} are satisfied.
The systematic study of the regimes when one or both of these criteria are violated, and when the effects of dissipation should be taken into account in the analysis of the photon transport through the system goes beyond the scope of this paper and will be presented elsewhere.

\section{Single-photon transport} \label{sec:transp}
\subsection{The effective Hamiltonian} \label{sec:ham_eff}
As was mentioned in Sec.~\ref{sec:princ}, to suppress the excitation exchange between the control qubit and the central resonator, the frequencies of the resonator and $\textstyle |g\rangle \leftrightarrow |e\rangle$ transition of the qutrit are strongly detuned from each other.
Provided that the condition
\begin{equation} \label{eq:disp_cond}
 |\lambda|\ll 1, \quad \lambda \equiv \frac{g_{ge}}{\omega_{ge} - \omegac},
\end{equation}
holds \cite{blais2004}, one can treat the interaction between the resonator and $\textstyle |g\rangle \leftrightarrow |e\rangle$ transition perturbatively and eliminate the interaction term $\textstyle \propto (a^\dag\sigma_{ge} + \sigma_{eg}a)$ in the Hamiltonian~\eqref{eq:ham_sw} using the Schrieffer-Wolff transformation \cite{klimov2000,blais2004}:
\begin{equation} \label{eq:sw}
 \opH \rightarrow \opH' = \ee^{-\lambda\opS} \opH \ee^{\lambda\opS}, \quad \opS = a^\dag_0 \sigma_{ge} - \sigma_{eg} a_0.
\end{equation}
The Schrieffer-Wolff transformation~\eqref{eq:sw} allows us to decouple the qubit eigenspace from the resonator field up to the first order in $\lambda$, and then project into the low-energy subspace of the resonator field \cite{bravyi2011}.

For deriving the transformed Hamiltonian $\textstyle \opH'$, we use Eq.~\eqref{eq:ham} along with the Baker-Campbell-Hausdorff relation
\begin{equation} \label{eq:bch_rel}
 \begin{split}
   \opH' & = \ee^{-\lambda \opS} \opH \ee^{\lambda \opS} \\
         & = \opH + \lambda[\opH,\opS] + \frac{\lambda^2}{2}[[\opH,\opS],\opS] + \ldots \, ,
 \end{split}
\end{equation}
where we keep only the terms contributing up to first order in $\textstyle \lambda$.
The form of terms $\textstyle \opH_\res$, $\textstyle \opH_{\res-\res}$, $\textstyle \opH_\mathrm{tl}$, and $\textstyle \opH_\mathrm{tl-r}$ are retained after applying the transformation, while the Hamiltonian of the resonator-qutrit system acquires the form $\textstyle \ee^{-\lambda \opS} \opH_\mathrm{s} \ee^{\lambda \opS} = \opH'_\mathrm{s}$:
\begin{equation} \label{eq:ham_sw1}
 \begin{split}
  \opH'_\mathrm{s} = & \, \hbar (\omegac + \chi \hat{Z}_{eg}) a^\dag_0 a_0 + \hbar (\omega_{ge} + \chi) \sigma_{ee} \\
    & \, + \hbar (\omega_{ge}+\omega_{ef})\sigma_{ff} + \hbar g_{ef}(a^\dag_0\sigma_{ef} + \sigma_{fe} a_0),
 \end{split}
\end{equation}
where $\textstyle \hat{Z}_{eg} = \sigma_{ee} - \sigma_{gg}$ and $\textstyle \chi = \lambda g_{ge}$.
In the transformed Hamiltonian $\textstyle \opH'$, we dropped $\textstyle \lambda J a^\dag_{\pm 1}\sigma_{ge}$, $\textstyle \lambda g_{ge} a^{\dag 2}_0 \sigma_{gf}$ and their conjugates, since these terms contribute in the order of $\textstyle \lambda^2$.

Since $\textstyle [\sigma_{ee}+\sigma_{ff},\opH']=0$, it is convenient to make a transformation
\begin{equation}
 \opH' \rightarrow \opH' - \hbar (\omega_{ge}+\chi)(\sigma_{ee} + \sigma_{ff}),
\end{equation}
which turns $\textstyle \opH'_\mathrm{s}$ into the Hamiltonian as follows
\begin{equation} \label{eq:ham_sw2}
 \opH'_\mathrm{s} = \hbar \bar{\omega}_\mathrm{c} a^\dag_0 a_0 + \hbar \omegaa \sigma_{ff} + \hbar g_{ef} (a^\dag_0 \sigma_{ef} + \sigma_{fe} a_0),
\end{equation}
where $\textstyle \bar{\omega}_\mathrm{c} = \omegac + \chi \hat{Z}_{eg}$ stands for the qubit--state-dependent frequency of the ``dressed" central resonator and $\textstyle \omegaa = \omega_{ef} - \chi$ denotes the frequency of the ``dressed" $\textstyle |e\rangle \leftrightarrow |f\rangle$ qutrit transition.
In what follows, for the description of the system dynamics, we use the Hamiltonian $\textstyle \opH'$ with $\textstyle \opH'_\mathrm{s}$ expressed by Eq.~\eqref{eq:ham_sw2}.

\subsection{Scattering dynamics} \label{sec:dyn}

The probability of finding the photon at time $\textstyle t$ in the output waveguide for the control qubit prepared in one of its eigenstates ($\textstyle |g\rangle$ or $|e\rangle$) is determined as
\begin{equation} \label{eq:trns}
 \mathcal{T}_q(t) = \int^\infty_0 \dd\omega \left|\langle \psi^q_{2,\omega}(t)|\Psi^q_\mathrm{in}\rangle\right|^2, \quad q\in\{g,e\},
\end{equation}
where
\begin{equation} \label{eq:def_psiq}
 \begin{split}
  & |\psi^q_{2,\omega}(t)\rangle = b^\dag_{2,\omega}(t)\vacqket, \\
  & \vacqket = |q\rangle |\varnothing\rangle_{\mathrm{w}1} |\varnothing\rangle_{\mathrm{w}2} \bigotimes_{n=-\Nres}^\Nres |\varnothing\rangle_{n\mathrm{r}}.
 \end{split}
\end{equation}
The state $\textstyle |\psi^q_{2,\omega}(t)\rangle$ corresponds to the state of the system hosting a single photon of frequency $\textstyle \omega$ propagating in the second (output) waveguide, the qubit residing in the excited state $\textstyle |q\rangle$ and void of excitations in the CRA and the first (input) waveguide.

In Eq.~\eqref{eq:def_psiq}, the state $\textstyle |\Psi^q_\mathrm{in}\rangle$ stands for the initial (at $\textstyle t=0$) state of the entire system
We set that initially the single-photon wave packet propagates in the input waveguide, while the CRA and the output waveguide contain no photons.
We assume that the number of thermal excitations $\textstyle n_\mathrm{th}$ in the system is negligible.
Superconducting cQED systems typically operate at frequencies $\textstyle \omega_\mathrm{s}/(2\pi) \sim \textrm{3--8 GHz}$ and the temperature of the cryogenic stage $\textstyle T_\mathrm{s} \sim \textrm{10--20 mK}$ \cite{krantz2019}. 
For that frequency range and setup working temperature, the upper estimate for the thermal photon number in the system is $\textstyle n_\mathrm{th}<10^{-3}$ assuming the Bose-Einstein distribution of thermal photons $\textstyle n_\mathrm{th} = [\exp(-\frac{\hbar \omega_\mathrm{s}}{k_\mathrm{B} T_\mathrm{s}}) - 1]^{-1}$, where $\textstyle k_\mathrm{B}$ is the Boltzmann constant.
Thus, the initial state of the system $\textstyle |\Psi^q_\mathrm{in}\rangle$ reads
\begin{equation} \label{eq:Psi_in}
 |\Psi^q_\mathrm{in}\rangle = |q\rangle|1_\xi\rangle_{\mathrm{w}1}|\varnothing\rangle_{\mathrm{w}2} \bigotimes_{n=-\Nres}^\Nres |\varnothing\rangle_{n\mathrm{r}},
\end{equation}
where $\textstyle |\varnothing\rangle_{\mathrm{w}j}$ is a state of the $\textstyle j$-th waveguide void of photons, and $\textstyle |\varnothing\rangle_{j\mathrm{r}}$ is a vacuum state of the $\textstyle n$-th resonator in the CRA.

\begin{figure*}[t!] 
	\centering
	\includegraphics{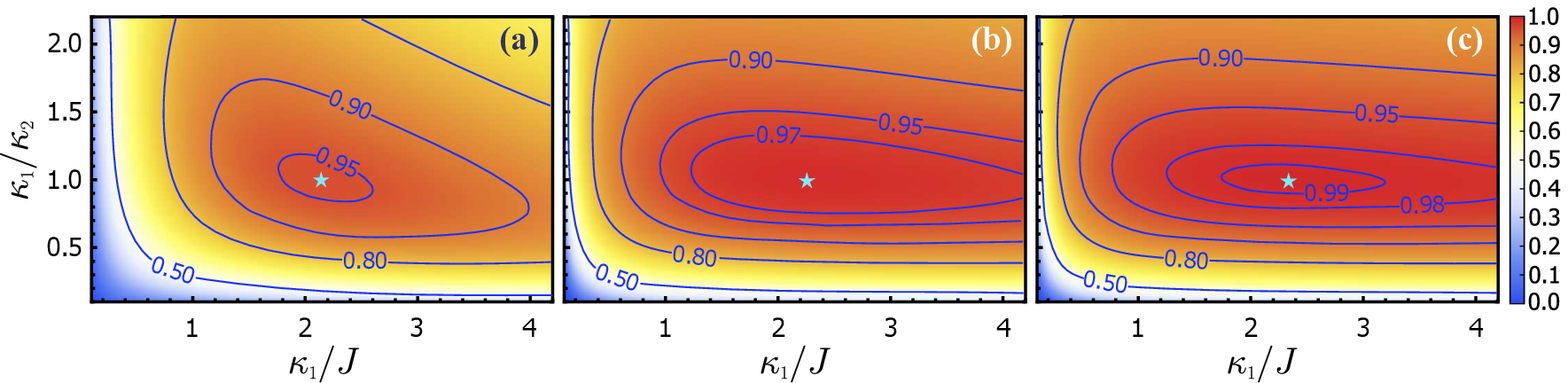}
	\caption{Dependence of the switching contrast on the interrelation between the waveguide-CRA photon exchange rates and the photon hopping rate in the CRA for different durations of the ingoing pulse: (a) $\textstyle \tau_\mathrm{p} = 0.1\,\mathrm{\mu s}$, (b) $\textstyle \tau_\mathrm{p} = 0.5\mathrm{\mu s}$, and
		(c) $\textstyle \tau_\mathrm{p} = 0.9\,\mathrm{\mu s}$.
		Stars mark the position of the maximal contrast: (a) $\textstyle \mathcal{C}_\mathrm{max}=0.956$, (b) $\textstyle \mathcal{C}_\mathrm{max}=0.989$, and (c) $\textstyle \mathcal{C}_\mathrm{max}=0.993$.
		The rest of the parameters are the following: $\textstyle J/(2\pi)=10\,\mathrm{MHz}$, $\textstyle g_{ef}/(2\pi) = 30\,\mathrm{MHz}$. \label{fig:fig_4}}
\end{figure*}

The state $\textstyle |1_\xi\rangle_{\mathrm{w}1}$ defined as
\begin{equation} \label{eq:def_1xi}
 |1_\xi\rangle_{\mathrm{w}1} \equiv \int^\infty_0\dd\omega \, \xi(\omega) b^\dag_{1,\omega}(0)|\varnothing\rangle_{\mathrm{w}1},
\end{equation}
stands for the state of the first (input) waveguide accommodating a single-photon wavepacket characterized by the \emph{spectral distribution function} \cite{rohde2007} denoted as $\textstyle \xi(\omega)$.
In the analysis, we assume that the ingoing wave packet is narrowband, and its spectrum is strongly localized near the central (carrier) frequency $\textstyle \omega_0$, i.e., $\textstyle \gamma_0\ll\omega_0$, where $\textstyle \gamma_0$ denotes the ingoing pulse bandwidth.

The probability of photon transmission $\textstyle \mathcal{T}_q(t)$ is governed by the evolution equation as follows (the derivation is given in Appendix~\ref{sec:appa2}):
\begin{equation} \label{eq:eq_TQ}
 \mathcal{T}_q(t) = \kappa_2 \int^t_0 \dd\tau \left|\vacqbra a_\Nres(\tau)|\Psi^q_\mathrm{in}\rangle\right|^2.
\end{equation}

Let us write down the equation of motion governing the matrix element $\textstyle \vacqbra a_\Nres(t)|\Psi^q_\mathrm{in}\rangle$ standing on the right-hand side of Eq.~\eqref{eq:eq_TQ}.
Using the Heisenberg equations for the CRA variables [see Eq.~\eqref{eq:eq_an} in Appendix~\ref{sec:appa1}], one obtains the evolution equation for $\textstyle A_n(t)$ as follows:
\begin{equation} \label{eq:eq_Aj}
 \ii \pdt A^q_n(t) = \omegar A^q_n(t) + J \left[A^q_{n-1}(t) + A^q_{n+1}(t)\right],
\end{equation}
where $\textstyle |n|\in[1, \Nres-1]$.
Here we introduced a notation $\textstyle A_n(t) = \vacqbra a_n(t)|\Psi^q_\mathrm{in}\rangle$.
For $\textstyle n=0$, one has
\begin{equation} \label{eq:eq_A0}
 \begin{split}
  \ii \pdt A^q_0(t) = & \, \left[\omegac + (2\eta_q-1)\chi\right] A^q_0(t) \\
  & \, + J\left[A^q_{-1}(t) + A^q_1(t)\right] + g_{ef} S^q_{ef}(t),
 \end{split}
\end{equation}
where $\textstyle \eta_q = |\langle e|q\rangle|^2$.
In Eq.~\eqref{eq:eq_A0}, we introduced a notation $\textstyle S^q_{ef}(t) = \vacqbra \sigma_{ef}(t)|\Psi^q_\mathrm{in}\rangle$.
Using the Heisenberg equation~\eqref{eq:eq_Sef} for the operator $\textstyle \sigma_{ef}$, one derives the equation of motion governing $\textstyle S^q_{ef}(t)$ as follows:
\begin{equation} \label{eq:eq_S}
  \ii \pdt S^q_{ef}(t) = \omegaa S^q_{ef}(t) + \eta_q g_{ef} \, A^q_0(t).
\end{equation}
Finally, for $\textstyle n=\pm\Nres$, one arrives at the following evolution equations (see derivation in Appendix~\ref{sec:appa3}):
\begin{subequations} \label{eq:eqs_An}
 \begin{equation} \label{eq:eq_AN}
   \ii \pdt A^q_\Nres(t) = \left(\omegar - \ii\frac{\kappa_2}{2}\right) A^q_\Nres(t) + J A^q_{\Nres-1}(t),
 \end{equation}
 \begin{equation} \label{eq:eq_An}
  \begin{split}
   \ii \pdt A^q_{-\Nres}(t) = & \, \left(\omegar - \ii\frac{\kappa_1}{2}\right) A^q_{-\Nres}(t) + J A^q_{1-\Nres}(t) \\
   & \, + f_1(\omega_0) \Xi(t).
  \end{split}
 \end{equation}
\end{subequations}
Function $\textstyle \Xi(t)$ is defined as
\begin{equation} \label{eq:def_Xi}
 \Xi(t) \equiv \int^\infty_{-\infty} \dd\omega \ee^{-\ii\omega t} \xi(\omega) = \sqrt{2\pi}\varrho(-t),
\end{equation}
where $\textstyle \varrho(t) = (2\pi)^{-1/2}\int^\infty_{-\infty}\dd\omega\ee^{\ii\omega t}\xi(\omega)$ describes the time-domain probability density amplitude of the ingoing pulse.

For computations we model the spectral distribution of the ingoing pulse $\textstyle \xi(\omega)$ by the Lorentzian function 
\begin{equation} \label{eq:xi_lor}
 \xi(\omega) = \sqrt{\frac{1}{2\pi \tau_\mathrm{p}}} \left[(\omega-\omega_0) + \frac{\ii}{2\tau_\mathrm{p}}\right]^{-1},
\end{equation}
which corresponds to the decaying exponent profile of the time-domain probability density amplitude
\begin{equation}
   \varrho(\tau) = \frac{1}{\sqrt{\tau_\mathrm{p}}}\exp\left(\frac{\tau}{2\tau_\mathrm{p}}+\ii\omega_0\tau\right)\theta(-\tau),
\end{equation}
where $\textstyle \tau_\mathrm{p}=1/\gamma_0$ stands for the ingoing pulse duration and $\textstyle \theta(\tau)$ is the Heaviside step function.
For convenience, we assume that the front of the ingoing pulse, which initially propagates in the first waveguide, reaches the CRA terminal resonator at instant $\textstyle t=0$.

We solve the system of differential equations~\eqref{eq:eq_Aj}--\eqref{eq:eq_An} numerically using the \texttt{NDSolve} function of \textsc{Mathematica}.

\section{Switching contrast} \label{sec:res} 
As a measure of the efficiency of the considered single-photon switch, we use a quantity given by
\begin{equation} \label{eq:c_def}
 \cC = \mathcal{T}_g(t_\infty) - \mathcal{T}_e(t_\infty),
\end{equation}
which is referred to as a switching contrast by analogy with a measurement contrast employed for the characterization of the accuracy of qubit measurement \cite{govia2014,sok2020}.
In Eq.~\eqref{eq:c_def}, $\textstyle t_\infty$ is attributed to the time, when all scattering processes in the system are finished and the scattered photon propagates in one of the waveguide as a free excitation. It is determined by the criterion $\textstyle t_\infty\gg\tau_\mathrm{tvl}$, where $\textstyle \tau_\mathrm{tvl} = \tau_\mathrm{p} + N_\mathrm{res}/(2J)$ is a photon travel time through the CRA.
For computations, we set $t_\infty = 10\tau_\mathrm{p}$.

We tune the frequency of the ``bare" central resonator $\textstyle \omegac$ to satisfy the relation $\textstyle \omegac - \chi = \omegar$.
Thus, when one prepares the control qubit in its ground state $\textstyle |g\rangle$, the frequency of the ``dressed" central resonator $\textstyle \langle g|\bar{\omega}_\mathrm{c}|g\rangle = \omegac - \chi = \omegar$ matches the frequencies of the other resonators in the CRA.
In this case, the incident photon propagates through the chain of resonators with identical couplings $J$ and frequencies $\textstyle \omegar$ resulting in the maximal transmission.
Using that $\textstyle \chi = g^2_{ge}/(\omega_{ge}-\omegac)$ and $\textstyle g_{ef} = \sqrt{2}g_{ge}$, one arrives at the relation between $\textstyle \omegac$ and $\textstyle \omegar$ as follows
\begin{equation} \label{eq:omegac}
 \omegac = \frac{1}{2}\left[\omegar + \omega_{ge} - \sqrt{(\omegar - \omega_{ge})^2 - 2 g^2_{ef}} \right].
\end{equation}
The frequency of the $\textstyle |e\rangle \leftrightarrow |f\rangle$ transition is set in such a way that when the qubit is prepared in its excited state $\textstyle |e\rangle$ one has $\textstyle \langle e|\bar{\omega}_\mathrm{c}|e\rangle = \omegac + \chi = \omegaa$.
Thus, in this scenario, the qutrit transition $\textstyle |e\rangle \leftrightarrow |f\rangle$ is ``switched on" and its ``dressed" frequency coincides with that of the central resonator that gives rise to the DIR effect leading to photon reflection.
Recalling that $\omegaa = \omega_{ef} - \chi$ and using Eq.~\eqref{eq:omegac}, one obtains
\begin{equation} \label{eq:omegaef}
 \omega_{ef} = \omegar + \frac{3 g^2_{ef}}{\omegar - \omega_{ge} + \sqrt{(\omega_r - \omega_{ge})^2 - 2g^2_{ef}}}.
\end{equation}

Now, let us proceed to the analysis of the performance of the proposed single-photon switch scheme.
To satisfy the criterion~\eqref{eq:disp_cond} of the dispersive regime of interaction between the resonator and the $\textstyle |g\rangle\leftrightarrow|e\rangle$ qutrit transition, we keep $\textstyle \lambda<0.1$ for all computations unless stated otherwise.
The relative anharmonicity $\textstyle \alpha_\mathrm{rel} = (\omega_{ef} - \omega_{ge})/\omega_{ge}$ of energy levels of the typical transmon artificial atom is around $\textstyle -0.05$ \cite{koch2007}.
Thus, in all calculations we choose the setup parameters in such a way that the relative anharmonicity of the qutrit is
$\textstyle -0.06 \le \alpha_\mathrm{rel} \le -0.04$.

Calculations of the dependence of the switching contrast $\textstyle \mathcal{C}$ on the interrelation between the photon hopping rate $\textstyle J$ and the CRA-waveguides exchange rates $\textstyle \kappa_{1,2}$ shown in Fig.~\ref{fig:fig_4} demonstrate that the maximal contrast $\textstyle \mathcal{C}_\mathrm{max}$ (for given values of $\textstyle J$ and $\textstyle g_{ef}$) is achieved when the CRA is equally coupled to both waveguides, i.e., $\textstyle \kappa_1 = \kappa_2$.
In what follows, we consider only this (symmetric) configuration of the setup.
In this regard, from now on, we use a notation $\textstyle \kappa \equiv (\kappa_1=\kappa_2)$ for brevity.

\begin{figure}[b]
	\centering
	\includegraphics{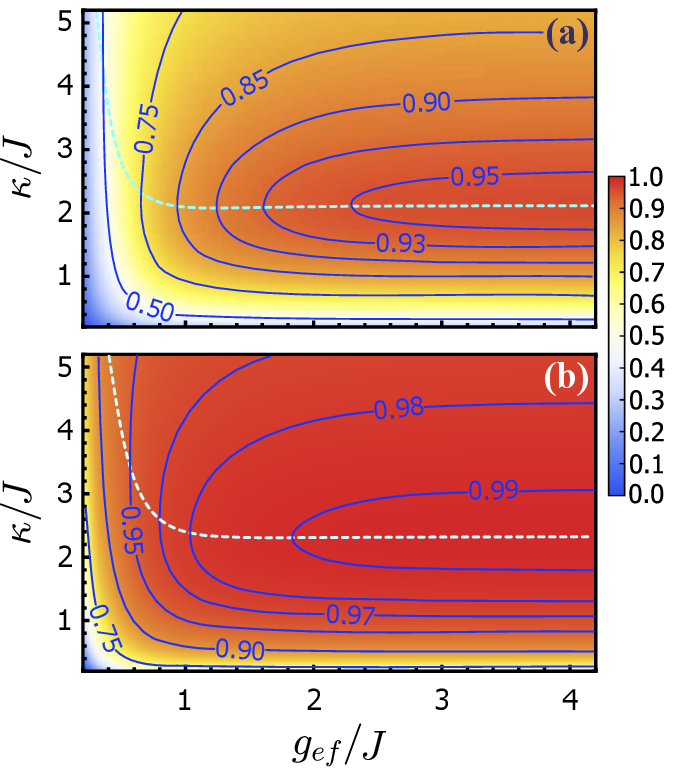}
	\caption{Dependence of the switching contrast on $\textstyle \kappa/J$ and $\textstyle g_{ef}/J$ for the ingoing pulse durations (a) $\textstyle \tau_\mathrm{p} = 0.1\mu\mathrm{s}$ and (b) $\textstyle \tau_\mathrm{p}=0.8\,\mathrm{\mu s}$.
		The dashed line marks the position of maximum $\textstyle \mathcal{C}_\mathrm{max}$ for given $\textstyle g_{ef}/J$.
		For calculations, we use the parameters as follows: $\textstyle J/(2\pi) = 10\,\mathrm{MHz}$, $\textstyle \omegar/(2\pi) = 7.0\,\mathrm{GHz}$,
		and $\textstyle \omega_{ef}/(2\pi)=7.36\,\mathrm{GHz}$.
		\label{fig:fig_5}}
\end{figure}

\begin{figure}[t] 
	\centering
	\includegraphics{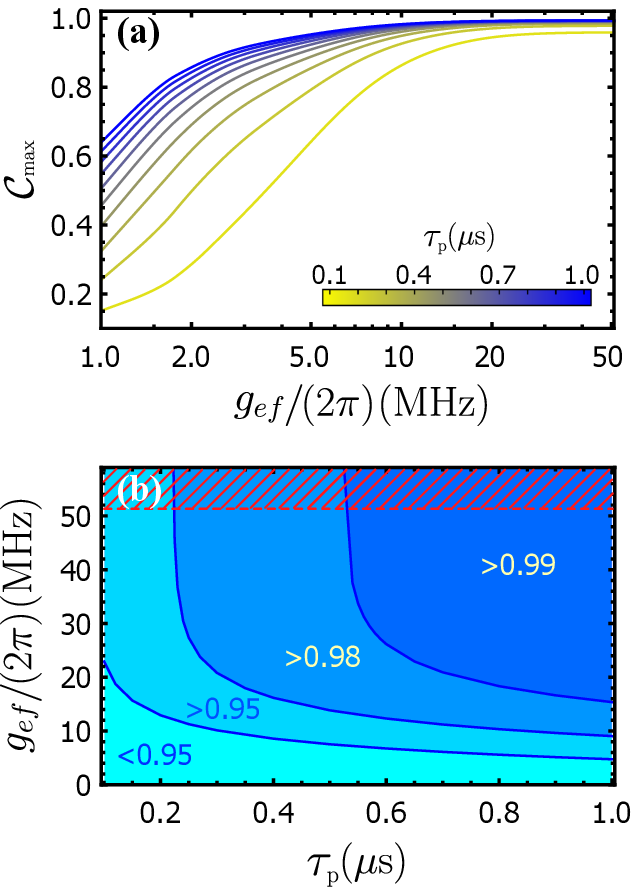}
	\caption{
		(a) Dependence of $\textstyle \mathcal{C}_\mathrm{max}$ on $\textstyle g_{ef}$ for the different ingoing pulse durations $\textstyle \tau_\mathrm{p}$ encoded by the color gradient.
		(b) The region plot demonstrating what switching contrast can be achieved for the given ingoing pulse duration and coupling $\textstyle g_{ef}$.
		The dashed region shows where $\textstyle \lambda\geq 0.1$ and the condition~\eqref{eq:disp_cond} of the dispersive coupling breaks down.
		The parameters used for calculations are the same as in Fig.~\ref{fig:fig_5}.
		\label{fig:fig_6}}
\end{figure}

Dependence of the switching contrast on the qutrit-resonator coupling and the ingoing pulse duration is shown in Figs.~\ref{fig:fig_5} and~\ref{fig:fig_6}a.
Computations reveal that the contrast improves with the increase of $\textstyle g_{ef}/J$.
The explanation is as follows.
Assume that one prepares the qubit in the excited state $\textstyle |e\rangle$.
Since we tune the ``bare" frequencies of the central resonator and the $\textstyle |e\rangle \leftrightarrow |f\rangle$ transition to obtain a resonance of the ``dressed" frequencies $\textstyle \omegac+\chi = \omegaa$, the single-excitation eigenfrequencies of the Jaynes-Cummings (JC) system composed of the central resonator and the 2LS formed by the qutrit levels $\textstyle |e\rangle$ and $|f\rangle$, are given by $\textstyle E^\pm_1  = \omegac \pm g_{ef}$.
Since we set $\textstyle \omegac-\chi=\omegar$, the eigenstates of the JC system are detuned from the frequencies of the neighbor resonators on $\textstyle (\delta_{\pm} = E^\pm_1 - \omegar) = \lambda g_{ge} \pm g_{ef}$.
For $\textstyle J<|\delta_{\pm}|$, photon hops from the resonator with index $n=-1$ on the central resonator start to be suppressed, which leads to photon reflection.
For the transmon, couplings $\textstyle g_{ge}$ and $\textstyle g_{ef}$ are of the same order of magnitude ($\textstyle g_{ef}\approx\sqrt{2}g_{ge}$ \cite{koch2007}).
Thus, the dominant contribution to the absolute value of the detuning $\textstyle |\delta_{\pm}|$ comes from $\textstyle g_{ef}$ since $|\lambda|\ll 1$ due to Eq.~\eqref{eq:disp_cond}.
Therefore, the increase of $\textstyle g_{ef}/J$ results in the larger probability of photon reflection and, thus, higher switching contrast.

Figure~\ref{fig:fig_6} demonstrates that the proposed single-photon switch can provide high contrasts for realistic values of the resonator-resonator and resonator-qutrit couplings and a wide range of the ingoing pulse durations.
For example, for $\textstyle J/(2\pi)=10\textrm{MHz}$ and $g_{ef}/(2\pi)\sim\textrm{30--50 MHz}$, the contrasts $\textstyle \mathcal{C}>0.95$ can be achieved for the ingoing pulses of duration $\textstyle \tau_\mathrm{p}>0.07\mathrm{\mu s}$, while for the longer pulses $\textstyle \tau_\mathrm{p}>0.55\,\mathrm{\mu s}$ the contrasts $\textstyle \mathcal{C}>0.99$ can be reached.
Results presented in Fig.~\ref{fig:fig_6} suggest that the switching contrast is limited by the incident pulse duration $\textstyle \tau_\mathrm{p}$ since shorter pulses exhibit a broader spectrum leading to a higher probability of unwanted reflection of a photon from the CRA, which reduces the switching contrast. 
To mitigate this effect, one can use higher values of $\textstyle \kappa$.
As follows from computations demonstrated in Fig.~\ref{fig:fig_5}, with the increase of $\kappa$, the resonator-resonator coupling $\textstyle J$ should be increased as well to avoid the reduction of the switching contrast.
However, larger $\textstyle J$ requires stronger qutrit-resonator couplings $\textstyle g_{ef}$ to keep the ratio $g_{ef}/J$.
Thus, for the fixed value of the resonator-qutrit coupling $\textstyle g_{ef}$, there should exist a combination of $\textstyle \kappa$ and $\textstyle J$, where the contrast reaches its maximal value $\textstyle C_\mathrm{max}$ for a given ingoing pulse duration $\textstyle \tau_\mathrm{p}$.
Due to the relation $\textstyle g_{ef} \approx \sqrt{2} g_{ge}$ holding for transmons \cite{koch2007}, the increase of $\textstyle g_{ef}$ entails the increase of $\textstyle g_{ge}$, requiring larger detuning between the qubit transition frequency and the central resonator frequency to ensure the condition~\eqref{eq:disp_cond} is fulfilled.
Larger detuning also implies the necessity of the stronger qutrit anharmonicity.
Moreover, the increase of detuning $\omegac-\omega_{ge}$ may, at some point, lead to a breakdown of the criterion~\eqref{eq:rwa_freqs}.
In this case, the RWA is not applicable, and the fast-oscillating terms in the qutrit-resonator Hamiltonian~\eqref{eq:ham_sw} should be taken into account.

To sum up the quantitative analysis of the performance of the proposed single-photon switch, we present Table~\ref{tab:tab1} aggregating several sets of setup parameters for reaching $\textstyle \mathcal{C} > 0.95$ for the sub-$\mathrm{\mu s}$ ingoing pulses.
For all parameter sets, we choose $\omegar/(2\pi) = 7.0\,\textrm{GHz}$, which lies in a range of typical values of CPWR frequency.
The transition frequency between the qubit levels is chosen $\omega_{ge}/(2\pi) = 7.36\,\textrm{GHz}$, ensuring that the relative anharmonicity is $\alpha_\mathrm{rel} \approx 0.05$.
Frequencies $\textstyle \omegac$ and $\textstyle \omega_{ef}$ are determined using Eqs.~\eqref{eq:omegac} and \eqref{eq:omegaef}.
Parameters $\kappa$ and $J$, presented in Table~\ref{tab:tab1}, provide the maximal switching contrast for the given values of $g_{ef}$ and $\tau_\mathrm{p}$.
Specific values of $g_{ef}$ and $\tau_\mathrm{p}$ are extracted from Fig.~\ref{fig:fig_6} to provide the switching contrasts around $0.95$, $0.98$, and $0.99$.
All parameters demonstrated in Table~\ref{tab:tab1} are achievable for the state-of-the-art superconducting cQED systems.

 \begin{figure}[t!] 
	\centering
	\includegraphics{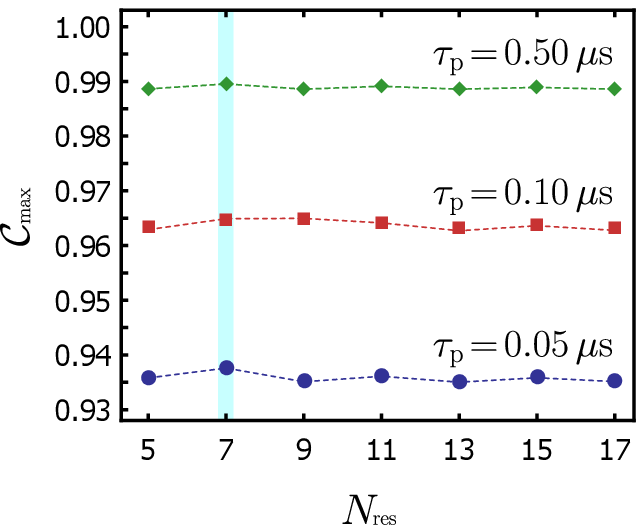}
	\caption{The maximal switching contrast $\textstyle \mathcal{C}_\mathrm{max}$, which can be reached for the given $\textstyle g_{ef}$ and $\textstyle \tau_\mathrm{p}$, as a function of the number of the resonators in the CRA $\textstyle N_\mathrm{res}$.
		The ingoing pulse durations $\textstyle \tau_\mathrm{p}$ are indicated near the corresponding point plots.
		For computation we used $\textstyle g_{ef}/(2\pi) = 40\,\textrm{MHz}$.
		The rest of the system parameters are the same as in Figs.~\ref{fig:fig_5} and \ref{fig:fig_6}.
		\label{fig:fig_7}}
\end{figure}

In Table~\ref{tab:tab1}, we also provide the values of parameter $\textstyle \varUpsilon$ (see definition in Appendix~\ref{sec:spec}) characterizing the degree of photon spectrum modification after passing the device.
Computations reveal that for the setup parameters providing high switching contrasts, one has $\textstyle \varUpsilon \ll 1$, indicating that the proposed single-photon switch introduces only a minor modification of the photon spectrum.

\begin{table*}[t]
	\caption{Realistic parameters of the setup to achieve high switching contrasts $\textstyle \mathcal{C}>0.95$ for the sub-$\mu\mathrm{s}$ ingoing pulses. \label{tab:tab1}}
	\begin{center}
		\begin{ruledtabular}
			\newcolumntype{1}{D{.}{.}{1}} 
			\newcolumntype{2}{D{.}{.}{2}} 
			\newcolumntype{3}{D{.}{.}{3}} 
			\newcolumntype{4}{D{.}{.}{4}} 
			\begin{tabular}{33332111234ccc}
				\hel{$\textstyle \omegar/2\pi$} &
				\he{$\textstyle \omegac/2\pi$} &
				\he{$\textstyle \omega_{ge}/2\pi$} &
				\he{$\textstyle \omega_{ef}/2\pi$} &
				\he{$\textstyle \alpha/2\pi$} &				
				\he{$\textstyle J/2\pi$} &
				\he{$\textstyle \kappa/2\pi$} &
				\he{$\textstyle g_{ef}/2\pi$} &
				\he{$\textstyle \tau_\mathrm{p}$} &
				\he{$\textstyle \mathcal{C}$} &
				\he{$\textstyle \varUpsilon$}
				\\
				\hel{(\textrm{GHz})} & \he{(\textrm{GHz})} & \he{(\textrm{GHz})} & \he{(\textrm{MHz})} & \he{(\textrm{GHz})}
				& \he{(\textrm{MHz})} & \he{(\textrm{MHz})} & \he{(\textrm{MHz})} & \he{($\mu\mathrm{s}$)} & & \\
				\hline
				7.000 & 7.004 & 7.360 & 7.011 & -349.48 & 20.0 & 45.0 & 50.0 & 0.06 & 0.952 & 0.0118 \\
				7.000 & 7.002 & 7.360 & 7.007 & -353.29 & 12.0 & 28.0 & 40.0 & 0.30 & 0.985 & 0.0044 \\
				7.000 & 7.001 & 7.360 & 7.004 & -356.24 & 10.5 & 24.0 & 30.0 & 0.60 & 0.991 & 0.0036
			\end{tabular}
		\end{ruledtabular}
	\end{center}
\end{table*}

All numerical results demonstrated in Figs.~\ref{fig:fig_4}--\ref{fig:fig_6} and Table~\ref{tab:tab1} were obtained for the CRA composed of $\textstyle N_\mathrm{res} = 7$ resonators.
Figure~\ref{fig:fig_7} demonstrates the dependence of the maximal switching contrast $\textstyle \mathcal{C}_\mathrm{max}$, which can be achieved for the given pulse duration $\textstyle \tau_\mathrm{p}$ and coupling $\textstyle g_{ef}$, on the number of resonators in the CRA $N_\mathrm{res}$.
Computations performed for $\textstyle N_\mathrm{res} \in [5\ldots 17]$ reveal that the dependence of $\textstyle \mathcal{C}_\mathrm{max}$ on $N_\mathrm{res}$ is minor.
The system with $N_\mathrm{res}=5$ provides slightly lower contrasts than $N_\mathrm{res} = 7$ for all $\tau_\mathrm{p}$.
For all $\textstyle N_\mathrm{res}$ and $\tau_\mathrm{p}$ used in Fig.~\ref{fig:fig_7}, we obtained $\varUpsilon < 0.015$.
Thus, $\textstyle N_\mathrm{res} = 7$ in the CRA already suffices for the efficient operation of the proposed single-photon switch.
The low number of required resonators may be beneficial for the realization of more complex systems requiring multiple single-photon switches.

\section {Discussion and Summary} \label{sec:concl}
\begin{figure}[b!] 
	\centering
	\includegraphics{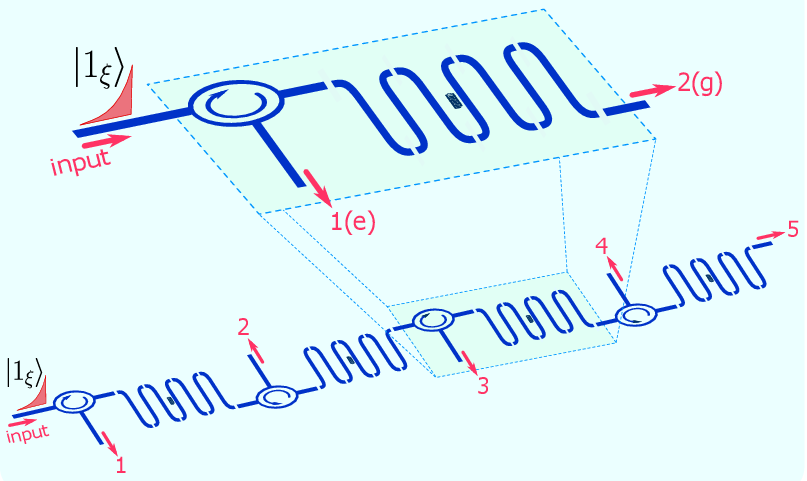}
	\caption{Scheme of a five-port single-photon router composed of a series of two-port routers.
		A two-port router is built by embedding a circulator into the input waveguide of a proposed single-photon switch.
		When the control qubit is prepared in the ground state $\textstyle |g\rangle$, the photon is routed to the output waveguide (port 2).
		The photon is routed to port 1 when one prepares the qubit in the excited state $\textstyle |e\rangle$. 
		\label{fig:fig_8}}
\end{figure}

  Having analyzed the performance of the proposed single-photon switch, let us discuss its possible applications.
  By inserting a circulator into the first (input) waveguide of the switch, one can implement a two-port quantum router.
  A nonreciprocal element (circulator) is required for the separation of the input and reflected signals into the different channels.
  Since in this scheme, both the signal and control information are quantum, one can regard the considered router as genuinely quantum \cite{lemr2013}.
  The multi-port routing can be achieved by connecting a number of those two-port single-photon routers in a cascade configuration as proposed, e.g., in Refs.~\cite{hoi2011,zhu2019}. 
  The scheme of this multi-port router is shown in Fig.~\ref{fig:fig_8}.
  Recent advances in the demonstration of on-chip microwave circulators \cite{barz2017,chapman2017} paves the way to entirely on-chip realization of the multi-port quantum router for microwave photons.

  Besides the multi-port single-photon router, one can harness the proposed switch to implement the high-fidelity readout of superconducting artificial atoms using the single-photon probe pulses.
  As it was pointed out in Ref.~\cite{sok2020}, the use of the single-photon probe allows one to avoid the readout errors arising from the nonorthogonality of the probe state, which is always the case for the coherent-state readout pulses.
  The high-efficient on-demand sources of microwave single-photon pulses are readily available \cite{forndiaz2017,peng2016,*yuzhou2020}.  
  Moreover, one can employ a detector of itinerant photons \cite{sath2016} attached to the output waveguide of the switch to provide a ``click" for a particular state of the qubit \cite{sok2020}.
  When one prepares the qubit in the ground state, the probe photon is transmitted through the switch to the output waveguide and the photodetector clicks.
  Conversely, when the qubit is prepared in the excited state, the switch is reflective, and the probe photon can not reach the detector.
  In this case, the latter gives no click.
  Promising theoretical proposals \cite{sath2014,koshino2015,*koshino2016,royer2018,maltz2020,iakupov2020,sokolov2020}, as well as recent experimental demonstrations \cite{inomata2016,kono2018,besse2018,lesc2020,dass2020} of itinerant microwave photon detectors, allow us to be optimistic about the perspectives of the near-term realization of the scheme for the superconducting qubit readout outlined above.

  To summarize, we have proposed a scheme of an efficient single-photon switch and examined its performance in detail.
  The possible superconducting cQED realization of the considered single-photon switch was outlined.
  We have demonstrated that parameters of the setup required for achieving high switching contrasts are feasible for the state-of-the-art superconducting cQED devices.
  A few applications of the proposed switch, namely, a multi-port quantum router and a scheme for a single-photon readout of a qubit state, were considered as well.

  Study of the effects of dissipation and system inhomogeneities (e.g., random variations of frequencies and couplings) on the efficiency of the considered switching scheme may be of interest.
  Also, assessing the efficiency of the proposed switch in the regime of the multi-photon input constitutes a possible direction for follow-up research.

\begin{acknowledgements}
 The author is grateful to Andrii Sokolov and Oleksandr Chumak for fruitful discussions.
 This work was partially supported by the National Academy of Sciences of Ukraine Grant for Young Scientists Research Groups (Grant No. 0120U100155).
\end{acknowledgements}

\appendix
\section{Derivation of equations of motion} \label{sec:appa}
\subsection{Heisenberg equations} \label{sec:appa1}
The effective Hamiltonian $\textstyle \opH'$ generates the following Heisenberg equations for the CRA variables:
\begin{equation} \label{eq:eq_an}
   \ii \pdt a_n = \omegar a_n + J \left(a_{n-1}+a_{n+1}\right).
\end{equation}
For the central resonator variable $\textstyle a_0$, one has
\begin{equation} \label{eq:eq_a0}
 \ii \pdt a_0 = (\omegac+\chi\hat{Z}_{eg}) a_0 + g_{ef} \sigma_{ef} + J (a_{-1} + a_1).
\end{equation}
The Heisenberg equations for the annihilation operators of a photon in the terminal resonators read as
\begin{subequations} \label{eq:eq_aN}
	\begin{equation}
	 \ii \pdt a_{\Nres} = \omegar a_\Nres + J a_{\Nres-1} + \int^\infty_0 \dd\omega f_{2}(\omega) b_{2,\omega},
	\end{equation}
	\begin{equation}
	 \ii \pdt a_{-\Nres} = \omegar a_{-\Nres} + J a_{1-\Nres} + \int^\infty_0 \dd\omega f_{1}(\omega) b_{1,\omega}.
	\end{equation}
\end{subequations}
The waveguides variables $\textstyle b_{1,\omega}$ and $\textstyle b_{2,\omega}$ obey the equations of motion
\begin{subequations} \label{eq:eq_tl}
	\begin{equation}
	 \ii \pdt b_{1,\omega} = \omega b_{1,\omega} + f_1(\omega) a_{-\Nres},
	\end{equation}
	and
	\begin{equation}
	\ii \pdt b_{2,\omega} = \omega b_{2,\omega} + f_2(\omega) a_\Nres,
	\end{equation}
\end{subequations}
The formal solutions of these equation read
\begin{subequations} \label{eq:eq_tl_sln}
	\begin{align}
 	 b_{1,\omega}(t) & = \tilde{b}_{1,\omega}(t) - \ii f_1(\omega) \int^t_0 \dd\tau \ee^{-\ii\omega(t-\tau)} a_{\Nres}(\tau), \\
	 b_{2,\omega}(t) & = \tilde{b}_{2,\omega}(t) - \ii f_2(\omega) \int^t_0 \dd\tau \ee^{-\ii\omega(t-\tau)} a_{-\Nres}(\tau),
	\end{align}
\end{subequations}
where $\textstyle \tilde{b}_{j,\omega}(t) = b_{j,\omega}(0)\ee^{-\ii\omega t}$ denotes the annihilation operator of a free-propagating photon in the $\textstyle j$-th waveguide.

Let us evaluate the integrals
\begin{gather*}
 \mathcal{I}_{j}(t) = \int^\infty_0 \dd\omega f_j(\omega) b_{j,\omega}(t), \quad j\in\{1,2\},
\end{gather*}
standing in Eqs.~\eqref{eq:eq_aN}.
Using Eqs.~\eqref{eq:eq_tl_sln}, one obtains
\begin{equation}
 \begin{split}
   \mathcal{I}_2(t) = & \, \int^\infty_0 \dd\omega f_2(\omega) \tilde{b}_{2,\omega}(t) \\ & \, - \ii \int^t_0 \dd\tau \int^\infty_0 \dd\omega f^2_2(\omega) \ee^{-\ii\omega (t-\tau)} a_{\Nres}(\tau),
 \end{split}
\end{equation}
Consider the second term on the right-hand side of the above equation.
It follows from Eq.~\eqref{eq:eq_An}, that one can write $\textstyle a_N(t) = \mathsf{a}_N(t)\ee^{-\ii\omegar t}$, where $\mathsf{a}_N(t)$ represents a slowly-varying component of the operator $\textstyle a_N(t)$.
Due to integration over $\textstyle \tau$, only the narrow region of frequencies in the vicinity of $\textstyle \omegar$ gives the dominant contribution to the integral. Thus, one can assume $\textstyle f_2(\omega)\approx f_2(\omegar)$. The lower boundary of integration over $\textstyle \omega$ can be extended to $\textstyle -\infty$.
Using these approximations along with the property $\textstyle \int^\infty_{-\infty}\dd\omega \ee^{-\ii\omega(t-\tau)} = 2\pi\delta(t-\tau)$, one obtains
\begin{equation} \label{eq:int_frfl}
  \mathcal{I}_{2} = \tilde{\mathcal{I}}_2 - \frac{\kappa_2}{2} a_{\Nres}, \quad \mathcal{I}_{1} = \tilde{\mathcal{I}}_1 - \frac{\kappa_1}{2} a_{-\Nres},
\end{equation}
where $\textstyle \kappa_{j} = 2\pi f^2_{j}(\omegar)$ and $\textstyle  \tilde{\mathcal{I}}_j$ is defined as
\begin{equation} \label{eq:def_I0}
 \tilde{\mathcal{I}}_j(t) \equiv \int^\infty_0 \dd\omega f_j(\omega) \tilde{b}_{j,\omega}(t).
\end{equation}
For evaluation of $\textstyle \mathcal{I}_{1}(t)$, we employed the analogous reasons as those used for evaluation of $\textstyle \mathcal{I}_{2}(t)$.
Finally, substituting Eq.~\eqref{eq:int_frfl} into Eqs.~\eqref{eq:eq_aN}, one arrives at the result
\begin{subequations} \label{eq:eq_aN1}
	\begin{align}
	\ii \pdt a_\Nres & = \left(\omegar - \ii\frac{\kappa_2}{2}\right)a_\Nres + J a_{\Nres-1} + \tilde{\mathcal{I}}_2, \\
	\ii \pdt a_{-\Nres} & = \left(\omegar - \ii\frac{\kappa_1}{2}\right)a_{-\Nres} + J a_{1-\Nres} + \tilde{\mathcal{I}}_1.
	\end{align} 
\end{subequations}
It follows from the above equations that parameter $\textstyle \kappa_{j}$ stands for the rate of the photon exchange between the CRA and the $j$-th waveguide.

Using the Hamiltonian~\eqref{eq:ham_sw2} and the commutation relation~\eqref{eq:s_comm}, one obtains
\begin{equation} \label{eq:eq_Sef}
 \ii \pdt \sigma_{ef} = (\omegaa-\chi a^\dag_0 a_0)\sigma_{ef} - g_{ef} \hat{Z}_{fe} a_0,
\end{equation}
where $\textstyle \hat{Z}_{fe} = \sigma_{ff} - \sigma_{ee}$.

\begin{widetext}
\subsection{Equation of motion for $\textstyle \mathcal{T}_q(t)$} \label{sec:appa2}
Let us derive the evolution equation for the transmission probability $\textstyle \mathcal{T}_q(t)$ given by Eq.~\eqref{eq:trns}.
Using Eq.~\eqref{eq:def_psiq} along with Eq.~\eqref{eq:eq_tl_sln}, one obtains
 \begin{equation} \label{eq:eqTQ1}
 \begin{split}
  \mathcal{T}_q(t) & = \int^\infty_0 \dd\omega \left|\vacqbra b_{2,\omega}(t)|\Psi^q_\mathrm{in}\rangle\right|^2 \\
  & = \int^\infty_0 \dd\omega \int^t_0 \dd \tau \int^t_0 \dd\tau' f^2_2(\omega) \ee^{\ii \omega (\tau-\tau')}
  \, \langle \Psi^q_\mathrm{in}|a^\dag_\Nres(\tau')\vacqket\vacqbra a_\Nres(\tau)|\Psi^q_\mathrm{in}\rangle,
 \end{split}
 \end{equation}
where we used that $\textstyle \tilde{b}_{2,\omega}(t)|\Psi^q_\mathrm{in}\rangle = 0$, which follows from Eqs.~\eqref{eq:Psi_in} and \eqref{eq:def_1xi}.
Next, using the similar consideration that led us to Eqs.~\eqref{eq:int_frfl}, we extend the lower limit of integration over photon frequencies to $\textstyle -\infty$ and make an approximation $\textstyle f_2(\omega) \approx f_2(\omegar)$.
Now, integration over $\textstyle \omega$ gives $\textstyle 2\pi\delta(\tau-\tau')$ leading to Eq.~\eqref{eq:eq_TQ}
\end{widetext}

\subsection{Derivation of Eqs.~\eqref{eq:eqs_An}} \label{sec:appa3}
 Using Eqs.~\eqref{eq:eq_aN} and \eqref{eq:def_I0}, one derives the equation of motion for $\textstyle A^q_{\pm\Nres}(t) = \vacqbra a_{\pm\Nres}(t)|\Psi^q_\mathrm{in}\rangle$ as follows
 \begin{subequations}
	\begin{equation} \label{eq:eq_An2}
	\begin{split}
	\ii\pdt A^q_{\Nres}(t) = & \, \left(\omegar-\ii\frac{\kappa_2}{2}\right)A^q_{\Nres}(t) + J A^q_{\Nres-1}(t) \\
	& \, + \int^\infty_0 \dd\omega \ee^{-\ii\omega t} f_2(\omega) \vacqbra b_{2,\omega}(0)|\Psi^q_\mathrm{in}\rangle.
	\end{split}
	\end{equation}
	\begin{equation} \label{eq:eq_AN2}
	\begin{split}
	\ii\pdt A^q_{-\Nres}(t) = & \, \left(\omegar-\ii\frac{\kappa_1}{2}\right)A^q_{-\Nres}(t) + J A^q_{1-\Nres}(t) \\
	& \, + \int^\infty_0 \dd\omega \ee^{-\ii\omega t} f_1(\omega) \vacqbra b_{1,\omega}(0)|\Psi^q_\mathrm{in}\rangle.
	\end{split}
	\end{equation}
\end{subequations}
 Let us consider the last terms on the right-hand side of the above equations.
 Employing Eq.~\eqref{eq:Psi_in} in Eq.~\eqref{eq:eq_An2}, one obtains $\textstyle \vacqbra b_{2,\omega}(0)|\Psi^q_\mathrm{in}\rangle = 0$, which immediately leads to Eq.~\eqref{eq:eq_AN}.
 In Eq.~\eqref{eq:eq_AN2}, one has $\textstyle \textstyle \vacqbra b_{1,\omega}(0)|\Psi^q_\mathrm{in}\rangle = \xi(\omega)$ resulting in
 \begin{equation*}
  \int^\infty_0 \dd\omega \ee^{-\ii\omega t} f_1(\omega) \xi(\omega) \approx f_1(\omega_0) \int^\infty_{-\infty} \dd\omega \ee^{-\ii\omega t} \xi(\omega),
 \end{equation*}
 where the above approximation is obtained using that the ingoing wave packet is narrowband $\gamma_0\ll\omega_0$.
 Thus, following the lines of derivation of Eqs.~\eqref{eq:eq_aN1}, one can set $\textstyle f_1(\omega)\approx f_1(\omega_0)$ and extend the lower limit of integration on $\textstyle \omega$ to $\textstyle -\infty$.
 Combining this result with Eq.~\eqref{eq:eq_AN2} and recalling the definition of $\textstyle \Xi(t)$ given by Eq.~\eqref{eq:def_Xi}, one arrives at Eq.~\eqref{eq:eq_An}.
\newpage
\section{Photon spectrum modification} \label{sec:spec}
As a measure of a modification of the outgoing (transmitted or reflected) photon spectrum, we introduce a parameter 
\begin{equation}
 \varUpsilon = \mathrm{max}\left\{\varUpsilon_g, \varUpsilon_e\right\},
\end{equation}
where $\textstyle \varUpsilon_q$ characterizes the modification of the outgoing photon spectrum, provided that the control qubit is prepared in the eigenstate $\textstyle |q\rangle$.
We define $\textstyle \varUpsilon_q$ as
\begin{equation} \label{eq:Y_q}
 \varUpsilon_q = 1 - \frac{\int^\infty_0 \dd\omega \, S^q_\mathrm{out}(\omega) S_\mathrm{in}(\omega)}{\int^\infty_0 \dd\omega \, S^2_\mathrm{in}(\omega)},
\end{equation}
where $\textstyle S_\mathrm{in}(\omega)=\left|\xi_\omega\right|^2$ is the spectrum of the ingoing single-photon wave packet.
In Eq.~\eqref{eq:Y_q}, $S^q_\mathrm{out}(\omega)$ stands for the spectrum of the outgoing wave packet, provided the qubit is prepared in the state $|q\rangle$.
If the qubit is prepared in its ground state $|g\rangle$, one expects transmission of the photon.
In this case, the photon spectrum is determined as
 \begin{equation}
  S^g_\mathrm{out}(\omega) = \left|\langle \varnothing_g|b_{2,\omega}(t_\infty)|\Psi^g_\mathrm{in}\rangle\right|^2.
 \end{equation}
If one prepares the qubit in the excited state $|e\rangle$, the photon is likely to be reflected.
Then, the spectrum of the reflected photon is given by
 \begin{equation}
 	S^e_\mathrm{out}(\omega) = \left|\langle \varnothing_e|b_{1,\omega}(t_\infty)|\Psi^e_\mathrm{in}\rangle\right|^2.
 \end{equation}
In the ideal case, when the photon spectrum is preserved after traversing the switch for both states of the qubit $\textstyle S_\mathrm{in}(\omega) = S^{g,e}_\mathrm{out}(\omega)$, one has $\varUpsilon = 0$. As a criterion of an acceptable operation of the switch, we take $\textstyle \varUpsilon \ll 1$.
\bibliography{bibliography}
\end{document}